\documentclass[reprint,aip,prl,amsmath,amssymb,reprint,twocolumn]{revtex4-2}

\usepackage{graphicx}                                   
\usepackage{dcolumn}                                    
\usepackage{bm}                                         
\usepackage{chemformula} 
\usepackage[utf8x]{inputenc} 
\usepackage[T1]{fontenc} 
\usepackage{mathrsfs}
\usepackage{braket}
\usepackage{amssymb}
\usepackage{appendix}
\usepackage{hyperref}
\usepackage{color}
\usepackage[normalem]{ulem}
\usepackage{dcolumn}
\usepackage{adjustbox}
\usepackage{tabularx}
\usepackage{tikz}
\usepackage{physics}
\usepackage{program}
\usepackage{graphicx}
\usepackage{hyperref}
\usetikzlibrary{arrows,shapes,positioning,shadows,trees}

\newcommand{\nl}{\nonumber \\}

\draft 

\begin{document}

\title{Physics-Informed Neural Networks and Beyond: Enforcing Physical Constraints in Quantum Dissipative Dynamics} 
\author{Arif Ullah}   
\email{arif@ahu.edu.cn}
\author{Yu Huang}   
\author{Ming Yang}
\affiliation{School of Physics and Optoelectronic Engineering, Anhui University, Hefei, 230601, Anhui, China}  
\author{Pavlo O. Dral}
\email{dral@xmu.edu.cn}
\affiliation{State Key Laboratory of Physical Chemistry of Solid Surfaces, College of Chemistry and Chemical Engineering, Fujian Provincial Key Laboratory of Theoretical and Computational Chemistry, and Innovation Laboratory for Sciences and Technologies of Energy Materials of Fujian Province (IKKEM), Xiamen University, Xiamen, 361005, Fujian, China}
\affiliation{Institute of Physics, Faculty of Physics, Astronomy, and Informatics, Nicolaus Copernicus University in Toruń, ul. Grudziądzka 5, 87-100 Toruń, Poland}

\date{\today}

\begin{abstract}
Neural networks (NNs) accelerate simulations of quantum dissipative dynamics. Ensuring that these simulations adhere to fundamental physical laws is crucial, but has been largely ignored in the state-of-the-art NN approaches. We show that this may lead to implausible results measured by violation of the trace conservation. To recover the correct physical behavior, we develop physics-informed NNs (PINNs) that mitigate the violations to a good extend. Beyond that, we propose a novel uncertainty-aware approach that enforces perfect trace conservation by design, surpassing PINNs. 


\end{abstract}

\maketitle

 
\section{Introduction}

Open quantum systems are ubiquitous in nature and have versatile applications across various domains such as loss of coherence in quantum information,\cite{breuer2016colloquium} quantum memory,\cite{khodjasteh2013designing} quantum transport,\cite{cui2006quantum} proton tunnelling in DNA\cite{slocombe2022open} and energy transfer in photosynthetic systems.\cite{zerah2021photosynthetic} Being a multi-body problem, the exact characterization of open quantum systems is not feasible owing to exponential growth in Hilbert space dimension and a large number of environment degrees of freedom. However, the problem becomes more tractable by tracing out environment degrees of freedom $\mathbf{Tr}_{\rm E} (\, \cdot \,)$ or treating the environmentm\cite{wang2022quantum} and/or system within the classical phase space.\cite{jain2022pedagogical, meyer1979classical} To investigate open quantum systems, numerous approaches have been developed so far, spanning from entirely classical\cite{liu2021unified,runeson2019spin} to fully quantum methods.\cite{tanimura1989time, makarov1994path, han2020stochastic, ullah2020stochastic, su2023extended, yan2021efficient,chen2022simulation, gong2018quantum}
While each of these approaches has been successful in its own right, they are hampered by many limitations, such as the inability to account for quantum effects, or demanding significant computational resources arising from the need of employing a very small descretization step due to stability constraints. Furthermore, the comprehensive integration of environmental effects, especially in highly non-Markovian scenarios, contributes significantly to the computational overhead.

Neural networks (NNs) present an efficient approach to learn complex spatio-temporal dynamics in high-dimensional space. NNs and other machine learning (ML) methods have proven to be proficient at predicting future time evolution of quantum states as a function of historical dynamics.\cite{ullah2021speeding, rodriguez2022comparative, herrera2021convolutional, ullah2021speeding, zhang2023excited, wu2021forecasting, lin2022automatic, bandyopadhyay2018applications, yang2020applications,tang2022fewest}
In addition, NNs can directly predict the future quantum states as a function of time and/or simulation parameters.\cite{ullah2022predicting, ullah2022one, ge2023four, akimov2021extending, lin2022trajectory}

However, a crucial aspect of quantum simulations is adherence to fundamental physical principles. In simulating open quantum systems, it is essential for an approach to uphold the core physical principle of conserving the trace (the sum of probabilities for all possible states) of the reduced density matrix (RDM $\tilde{\boldsymbol{\rho}}_{\rm S}$), which should always be equal to 1, i.e., $\mathbf{Tr}_{\rm S}(\tilde{\boldsymbol{\rho}}_{\rm S}) = 1$, where $\mathbf{Tr}_{\rm S}$ represents trace over system degrees of freedom.

Despite the appeal of NNs, existing research on ML-based simulations of quantum dissipative dynamics has largely ignored trace conservation.\cite{ullah2021speeding, rodriguez2022comparative, herrera2021convolutional, ullah2021speeding, rodriguez2022comparative, herrera2021convolutional, ge2023four, zhang2023excited, wu2021forecasting, lin2022automatic, bandyopadhyay2018applications, yang2020applications,lin2022trajectory,tang2022fewest,ullah2022predicting, ullah2022one} To the best of our knowledge, only one study has mentioned, albeit in the context of a relatively simple system (spin-boson), that ML models, given sufficient data, were able to implicitly learn trace conservation to a reasonable degree.\cite{herrera2021convolutional} However, we cannot expect that it always holds, especially in much more complex situations and when it is difficult to obtain ample amount of data for implicit learning of the trace conservation. In general, the ML models can implicitly learn physical laws from the data but if left unchecked (unconstrained) or applied for situations too far from the training data, they can also spectacularly fail.

Physics-Informed Neural Networks (PINNs), introduced in 2017,\cite{raissi2017physics-a, raissi2017physics-b, raissi2019physics} present a promising solution to this problem.\cite{karniadakis2021physics, cuomo2022scientific} By incorporating physical constraints directly into the neural network architecture, PINNs ensure that the model's predictions adhere to underlying physical laws. This approach has been successfully applied across various fields, including fluid dynamics,\cite{cai2021physics-a, cai2021physics-b} seismic inversions in 2D acoustic media,\cite{rasht2022physics} chemical simulations,\cite{hou2024physics} quantum dynamics,\cite{zhang2024artificial} and electronic structure calculations.\cite{martinetto2024inverting}

In this paper, we explore whether NNs inherently conserve trace and demonstrate that unconstrained models can lead to unphysical results due to trace violations. To address this, we develop physics-informed neural networks that significantly reduce trace conservation violations. However, we find that even with the integration of physical knowledge, physics-informed NNs alone are not sufficient. To ensure correct physical behavior, we introduce an uncertainty-aware hard constraint (U-aware HC) approach that enforces perfect trace conservation by design.

The subsequent sections of this paper are structured as follows. In the "Theory and Methodology" section, we establish the foundational theory of open quantum systems and detail the various NN models employed in our study, including physics-agnostic and unconstrained NNs. We highlight the trace violations by these models and introduce physics-informed NNs (PINNs). Additionally, we discuss the associated loss functions used for training these models and introduce the U-aware HC constraint for rigorous enforcement of physical constraints. Following that, in the "Results and Discussion" section, we present our findings, comparing the performance of our PINN approach and HC constraint against existing models. We discuss the effectiveness of these approaches in enforcing physical laws and achieving accurate simulations. Finally, in the "Concluding Remarks" section, we summarize our key findings, explore the broader implications of our study, and outline potential future research directions.

\section{Theory and Methodology} \label{sec:theory}
 Let us consider an open quantum system ($\rm S$) with $n$-number of states coupled to an outside environment ($\rm E$). The dynamics of the composite system ($\rm S + \rm E$) is governed by the Liouville--von Neumann equation ($\hbar=1$) 
\begin{equation}
    \dot{\boldsymbol{\rho}}(t) = -i[\mathbf{H}, \boldsymbol{\rho}(t)],
\end{equation}
where $\mathbf{H}$ and $\boldsymbol{\rho}(t)$ represent Hamiltonian and density matrix of the composite system, respectively. As the composite system is an isolated systems, the dynamics is unitary. Assuming the initial state of the system and environment is uncoupled (i.e., $\boldsymbol{\rho}(0) = \boldsymbol{\rho}_{\rm S}(0) \otimes \boldsymbol{\rho}_{\rm E}(0)$), the non-unitary reduced dynamics of the system can be extracted by taking a partial trace over environment degrees of freedom, i.e., 
\begin{equation}
    \tilde{\boldsymbol{\rho}}_{\rm S}(t) = \mathbf{Tr}_{\rm E} \left( \mathbf{U}(t,0) \boldsymbol{\rho}(0) \mathbf{U}^\dagger(t,0)\right),
\end{equation}
where $\tilde{\boldsymbol{\rho}}_{\rm s}(t)$ is the reduced density matrix (RDM) of the system at time $t$, $\mathbf{Tr}_{\rm E}$ is the partial trace over environment degrees of freedom, and $\mathbf{U}(t,0)\left(\mathbf{U}^\dagger(t,0)\right)$ is the forward (backward) propagation operator in time. While most real-world systems technically qualify as "open" due to their environment, the immense complexity arising from all possible environmental interactions (known as the curse of dimensionality) makes exact theoretical solutions impractical. In the following, we present a brief theory of two broadly studied pedagogical systems: the two-state SB model and the Fenna-Matthews-Olson Complex (FMO) complex.

\hspace{0.5pt}

\noindent\textit{SB model}:
The SB model describes the temporal evolution of a qubit system (two-state system) interacting with an environmental bath comprising multiple independent harmonic oscillators. The system's total Hamiltonian, expressed in the basis of the excited (\(\ket{e}\)) and ground (\(\ket{g}\)) states, is given by:
\begin{equation}\label{eq:sb}
    \mathbf{H} = \epsilon \boldsymbol{\sigma}_z + \Delta \boldsymbol{\sigma}_x + \sum_{k} \omega_k \mathbf{b}_k^\dagger \mathbf{b}_k + \boldsymbol{\sigma}_z \sum_{k} c_k (\mathbf{b}_k^\dagger + \mathbf{b}_k),
\end{equation}
where \(\boldsymbol{\sigma}_z\) and \(\boldsymbol{\sigma}_x\) are the Pauli matrices, \(\epsilon\) is the energy bias of the qubit, and \(\Delta\) is the coupling strength between states. The environment's creation and annihilation operators for the \(k\)th mode are \(\mathbf{b}_k^\dagger\) and \(\mathbf{b}_k\), respectively, with \(\omega_k\) being the mode's frequency. The coupling strength between the system and the \(k\)th environmental mode is denoted by \(c_k\). The environmental influence on the system is characterized by an Ohmic spectral density function with a Drude-Lorentz cutoff:\cite{caldeira1983path} 
\begin{equation}\label{eq:drude}
    J(\omega) = 2 \lambda \frac{\gamma \omega}{\omega^2 + \gamma^2},
\end{equation}
with \(\lambda\) being the reorganization energy and \(\gamma\) representing the characteristic frequency or the reciprocal of the environmental relaxation time, \(\gamma = 1/\tau\).

\hspace{0.5pt}

\noindent\textit{FMO Complex}:
The FMO complex, a trimer in green sulfur bacteria, plays a crucial role in photosynthesis. Each monomer in the complex contains chlorophyll molecules that act as energy transfer sites, typically numbering seven or eight.\cite{am2011eighth} The energy transfer within an FMO monomer is described by the Frenkel exciton model Hamiltonian:\cite{ishizaki2009unified}  
\begin{align} \label{eq:fmo_hamil}
    \mathbf{\rm H} = &  \sum_{n=1}^{N} \ket{n}\epsilon_n \bra{n} + \sum_{n \neq m}^{N} \ket{n} J_{nm} \bra{m} \nl
    & + \sum_{n=1}^{N} \sum_{k=1} \left(\frac{1}{2} \mathbf{\rm P}_{k, n}^{2} + \frac{1}{2} \omega_{k, n}^{2} \mathbf{\rm Q}_{k, n}^{2}\right) \mathbf{I} \nl
    & - \sum_{n=1}^{N} \sum_{k=1} \ket{n} c_{k,n} \mathbf{\rm Q}_{k, n} \bra{n} + \sum_{n=1}^{N} \ket{n} \lambda_{n} \bra{n},
\end{align}
where \(N\) is the number of chlorophyll sites, \(\epsilon_n\) is the on-site energy, and \(J_{nm}\) is the coupling strength between sites \(n\) and \(m\). The environmental contribution is represented by \(P_{k,n}\) and \(Q_{k,n}\), the momentum and coordinate of the \(k\)th mode interacting with site \(n\), with \(\omega_{k,n}\) as the mode's frequency. The identity matrix \(\mathbf{I}\) ensures dimensional consistency. \(c_{k,n}\) is the coupling strength between the \(k\)th mode and site \(n\), and \(\lambda_n\) is the reorganization energy for site \(n\).

For our analysis, we utilize the same Ohmic spectral density function with a Drude-Lorentz cutoff as in Eq.~\eqref{eq:drude}, assuming a uniform spectral density across all sites.

\subsection{NN-accelerated quantum dissipative dynamics}

Within NN framework, learning the time evolution of $n$-dimensional RDM can be defined as learning a mapping function $\Psi: \mathbb{R}^{n \times k} \mapsto  \mathbb{R}^{n \times r}$ which takes a vector of descriptive variables, $\mathbf{x} \in \mathbb{R}^{n \times k}$ , and maps it to the corresponding target RDM, $\mathbf{y} \in \mathbb{R}^{n \times r}$. NN approaches for this task can be categorized into two main types: recursive and non-recursive, depending on how they handle the descriptor and inference.

\textit{Recursive methods}: In recursive NN methodologies,\cite{ullah2021speeding, rodriguez2022comparative, herrera2021convolutional} a mapping function denoted as $\Psi_{\rm rec}$, is employed to predict future RDMs based on their past history. This approach mimics traditional quantum dynamics, where the evolution at any given time explicitly depends on the current state and implicitly on the past states. Mathematically, a recursive method can be described as:
\begin{align}
\Psi_{\text{Rec}}: & \{\mathbb{R}^{n \times r}\} \rightarrow \mathbb{R}^{r} \quad \text{such that} \nl & \quad \Psi_{\text{rec}} \left[ \dots, \tilde{\boldsymbol{\rho}}_{\rm S}(t_{m-1}), \tilde{\boldsymbol{\rho}}_{\rm S}(t_m) \right] = \tilde{\boldsymbol{\rho}}_{\rm S}(t_{m+1}),
\end{align}
where \(\{\boldsymbol{\cdot}\}\) represents a sequence containing the history of RDMs, denoted as \([..., \tilde{\boldsymbol{\rho}}_{\rm S}(t_{m-1}),\tilde{\boldsymbol{\rho}}_{\rm S}(t_m)]\), with $n$ being the number of time steps and $r$ the dimensionality of the $\tilde{\boldsymbol{\rho}}_{\rm S}$. Recursive methods make predictions iteratively. The predicted RDM ($\tilde{\boldsymbol{\rho_{\rm S}}}$) at time t is added to the history, and the oldest one is discarded to maintain a fixed-size memory. This updated history becomes the new input for the next prediction.

\textit{Non-recursive methods}: 
Non-recursive methods, as seen in \citenum{ullah2022predicting, ullah2022one} learn the mapping function $\Psi$ as a function of simulation parameters and/or temporal information. The time-dependent non-recursive method as used in \citenum{ullah2022predicting}, establishes a mapping function ($\Psi_{\rm AIQD}$) between RDM and simulation parameters including time $t$. Mathematically:
\begin{align}
    \Psi_{\text{AIQD}}: & \mathbb{R}^{p} \rightarrow \mathbb{R}^{r} \nl &
\quad \text{such that} \quad \Psi_{\text{AIQD}}(t, \mathbf{p}) = \tilde{\boldsymbol{\rho}}_{\text{S}}(t).
\end{align}
where \(\mathbf{p}\) is a vector containing simulation parameters (e.g., temperature, frequency, coupling strength). This approach allows for parallel computation of all time steps since the prediction for each step does not depend on the output of the previous step. In time-independent non-recursive methodology,\cite{ullah2022one} the mapping function \(\Psi_{\text{OSTL}}\) predicts the entire trajectory of the RDM for a set of time steps \(t_1\) to \(t_k\) in one go:
\begin{align}
    \Psi_{\text{OSTL}}: &  \mathbb{R}^{p} \rightarrow \mathbb{R}^{k \times r}
\quad \text{such that} \nl & \quad \Psi_{\text{OSTL}}(\mathbf{p}) = [\tilde{\boldsymbol{\rho}}_{\rm s}(t_1), \tilde{\boldsymbol{\rho}}_{\rm s}(t_2), ..., \tilde{\boldsymbol{\rho}}_{\rm s}(t_k)].
\end{align}
where the descriptor includes only the simulation parameters.

\subsection{Limitations of existing NNs for open quantum systems: purely data-driven approaches}
In the NN framework, establishing the mapping function \(\Psi\) between the descriptor \(x\) and its target dynamics can be approached in two ways: purely data-driven or based on known physical laws and constraints. Unfortunately, current research, including our own work\cite{ullah2021speeding, ullah2022predicting, ullah2022one}, on machine learning (ML)-based simulations of open quantum systems, relies exclusively on data-driven approximations of the mapping function \(\Psi\). As a result, these models often fail to capture underlying physical laws, leading to non-physical RDMs that violate trace conservation.

 Here, we classify purely data-driven NNs into two categories: "physics-agnostic NNs" and "unconstrained NNs". Physics-agnostic NNs are models that are not exposed to the complete data and thus remain unaware of the underlying physical laws and constraints. Unconstrained NNs, in contrast, are exposed to the entire data but do not incorporate physical constraints in their loss functions. 

\begin{figure}[!thb]
    \centering
    \includegraphics[width=0.5\textwidth]{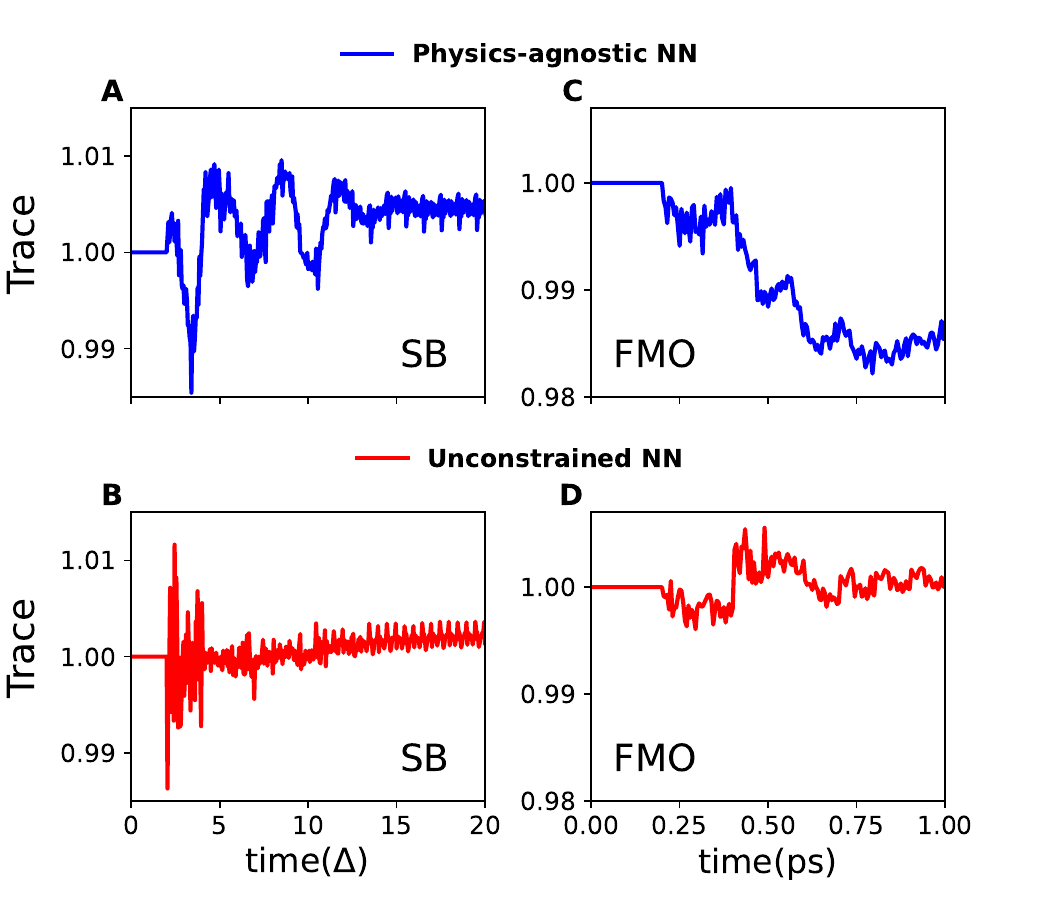}
    \caption{Trace violations in quantum dissipative dynamics using machine learning. Panels A and C in their respective order illustrate trace violations in a physics-agnostic scenario for a symmetric SB model and the 7-site FMO complex, where a CNN is trained for each state (site). Panels B and D demonstrate the improvement achieved with the unconstrained multi-output CNN for the same two systems. In all cases, an initial dynamics of length 0.2 (in the respective time units), with ideal trace conservation, serves as the seed for model predictions, derived from reference calculations. For the symmetric SB model, results are shown for an unseen dynamics with a characteristic frequency \(\gamma/\Delta = 9.0\), system-bath coupling \(\lambda/\Delta = 0.6\), and inverse temperature \(\beta\Delta = 1.0\). For the FMO complex, the initial excitation is considered on site-1, with parameters \(\gamma = 400~\text{cm}^{-1}\), \(\lambda = 40~\text{cm}^{-1}\), and temperature \(T = 90~\text{K}\). Further details on training and prediction can be found in the Results and Discussion section.}
    \label{fig:no-trace-conser}
\end{figure}

To emphasize on the issue of trace-violation by these data-driven NNs, we show their performance in Fig.~\ref{fig:no-trace-conser} with two examples: the relaxation dynamics within the SB model and the excitation energy transfer (EET) in the 7-site Fenna-Matthews-Olson (FMO) complex. As shown, these data-driven models fail to conserve the trace in both processes. In each case, we utilize convolutional neural networks (CNNs) and OSTL-based recursive dynamics propagation (Rec-OSTL)
\begin{align}\label{eq:ostl-rec}
\Psi_{\text{Rec-OSTL}}: & \{\mathbb{R}^{n \times r + p}\} \rightarrow \mathbb{R}^{k \times r} \quad \text{such that} \nl &  \Psi_{\text{Rec-OSTL}} \left[ \dots, \tilde{\boldsymbol{\rho}}_{\rm S}(t_{m-1}), \tilde{\boldsymbol{\rho}}_{\rm S}(t_m), \mathbf{p}\right] = \nl & [\tilde{\boldsymbol{\rho}}_{\rm s}(t_1), \tilde{\boldsymbol{\rho}}_{\rm s}(t_2), ..., \tilde{\boldsymbol{\rho}}_{\rm s}(t_k)]\, .
\end{align}
We use \texttt{MLQD} package\cite{ullah2024mlqd} and train the models on data from the \texttt{QD3SET-1} database\cite{ullah2023qd3set} (see Results and Discussion section for details). The training approach mirrors state-of-the-art methods reported previously.\cite{ullah2022one, rodriguez2022comparative}

In essence, for the physics-agnostic scenario (Fig.~\ref{fig:no-trace-conser}A and C), we train individual CNNs for each diagonal RDM element, employing a loss function that gauges the deviation of NN-predicted values $\bar{\tilde{\rho}}_{{\rm S},nn}$ from their reference counterparts $\tilde{\rho}_{{\rm S},nn}$:

\begin{equation}
    \mathcal{L}_{nn} =  \sum_{m=1}^M \left(\bar{\tilde{\rho}}_{{\rm S},nn,m} - \tilde{\rho}_{{\rm S},nn,m} \right)^2, 
\end{equation}
where $M$ is the number of training points and $m$ is the training point index.

As these models are not exposed to the dynamics of all states, they lack knowledge of trace conservation. We show that a much better solution is the unconstrained NN---a single, multi-output CNN designed to learn all RDM elements, incorporating a loss function that aggregates errors across all states (sites) (Fig.~\ref{fig:no-trace-conser}B and D):

\begin{equation}
    \label{eq:Lmulti}
    \mathcal{L}_\text{multi} =  \sum_{n=1}^N \mathcal{L}_{nn}. 
\end{equation}
However, despite being exposed to the dynamics of all states, this solution still exhibits minor but noticeable trace violations. It is important to note that trace violations can be reduced to some extent with additional training, as demonstrated in Fig. S1 of the Supporting Information. However, further improvement becomes limited as the model approaches the point of overfitting. Additionally, our observations indicate that the improvement in trace conservation with increasing memory time $t_m$ is somewhat unpredictable and does not follow a consistent trend. Despite this, there was a noticeable improvement in the accuracy of the dynamics predictions, as shown in Table S1.

\subsection{Our proposed solution: PINNs and beyond}
In the preceding subsection, we explored the shortcomings of the state-of-the-art purely data-driven NNs for simulating open quantum systems where they often struggle to enforce fundamental physical laws like trace conservation. This leads to inaccurate and non-physical results. To address this limitation, we first explore PINNs which integrate physical constraints into the loss function inspired by similar ideas in the literature.\cite{norambuena2024physics, muller2023exact} In our case, we include the additional loss term $\mathcal{L}_\text{Tr}$ penalizing the NN for the deviations from the trace conservation:

\begin{equation}
    \mathcal{L}_\text{PINN} =  \alpha \sum_{n=1}^N \mathcal{L}_{nn} + \eta \mathcal{L}_\text{Tr}, 
\end{equation}
where
\begin{equation}
    \mathcal{L}_\text{Tr} =  \sum_{m=1}^M \left( 1 - \sum_{n=1}^N \tilde{\rho}_{{\rm S},nn,m} \right)^2. 
\end{equation}

In these equations, we can tune $\mathcal{L}_{nn}$ and the deviations from trace conservation by weight factors $\alpha$ and $\eta$, respectively. Here we use $\alpha=2.0$ and $\eta=1.0$. Note that the unconstrained NN with the loss defined by Eq.~\eqref{eq:Lmulti} is a special case of the PINN with $\alpha = 1.0$ and $\eta=0$.

While PINNs significantly improve trace conservation compared to purely data-driven NNs, they can still exhibit minor violations (as we'll demonstrate later). This is because the physical constraints incorporated within the PINNs loss function are typically considered "soft." In simpler terms, PINNs are nudged towards satisfying the constraints during training, but they aren't strictly enforced.\cite{wang2021physics, norambuena2024physics} 

To overcome the limitations of PINNs, we propose a novel approach that enforces trace conservation by design. This approach utilizes an U-aware HC (uncertainty-aware hard-coded) constraint, guaranteeing strict adherence to physical laws during simulations. Unlike PINNs, the U-aware HC constraint operates outside of the loss function. This allows for a more direct and rigorous enforcement of the trace conservation law, rectifying potential violations during the simulation process.

The key idea is as follows: After making predictions with machine learning models, there will inevitably be a deviation from perfect trace conservation. We can calculate this residual deviation for each time step as:

\begin{equation}
    \Delta \text{Tr}(t) =  1 - \sum_{n=1}^N \tilde{\rho}_{{\rm S},nn}(t). 
\end{equation}

We can redistribute the residual deviations between each state as:

\begin{equation} \label{eq:trace-correct}
    \tilde{\rho}_{{\rm S}, nn}^\text{HC}(t) = \tilde{\rho}_{{\rm S}, nn}(t) + w_n(t) \Delta \text{Tr}(t).
\end{equation}

Here, we need to make such a choice for state-specific weighting factors $w_n$ that the trace is one. Also, it should be statistically motivated. Different states might be predicted with different uncertainty and for certain predictions we want smaller corrections (smaller weighting factors). Hence, we also need state-specific uncertainty quantification (UQ) of NN predictions. Similar problems were also faced in the prediction of partial atomic charges predicted by statistical models which do not necessarily add up to integer values: the suggested solution also was to redistribute the deviation from the correct total charge over atoms based on the UQ calculated as the disagreement between the models in ensemble~\cite{charge1,charge2}. This shows how very different research field can inspire the solutions in the unrelated field.

Here we introduce a new approach for UQ. We train an additional, auxiliary multi-output CNN with the same loss function as the main PINN but we shift the reference values by a prior $p_2$ (we assume that the main PINN model is trained with prior $p_1 = 0$). In other words, we train the CNN on $\tilde{\boldsymbol{\rho}}_{\rm S} + p_2 \mathbf{J}$ ($\mathbf{J}$ is a unit matrix with all elements 1) with the predictions given by:

\begin{equation} 
    \tilde{\rho}_{{\rm S}, nn}^\text{aux}(t) = \tilde{\rho}_{{\rm S}, nn}^\text{aux-NN}(t) - p_2\mathbf{J}.
\end{equation}

The UQ metric is given then as the absolute deviation of the $\tilde{\rho}_{{\rm S}, nn}^\text{aux}(t)$ from the main model predictions:

\begin{equation} \label{eq:uq_metric}
    D_{nn}(t) = \abs{\tilde{\rho}_{{\rm S}, nn}^\text{aux}(t) - \tilde{\rho}_{{\rm S}, nn}(t)}.
\end{equation}

The state-specific weighting factors $w_n$ we suggest to obtain as the normalized distances:

\begin{equation} \label{eq:weight}
    w_n(t) = \frac{D_{nn}(t)}{\sum_{n=1}^N D_{nn}(t)}.
\end{equation}

The implementation of Eq.~\eqref{eq:trace-correct} with the weighting factors defined with the Eq.~\eqref{eq:weight} ensures that 
$\mathbf{Tr}_{\rm S}\left(\tilde{\boldsymbol{\rho}}_{\rm S}^\text{HC}(t)\right)  = 1$. It's crucial to distinguish our proposed U-aware HC constraint-based approach from the conventional trace normalization technique, $\tilde{\rho}_{\rm S}/\mathbf{Tr}_{\rm S}\left(\tilde{\rho}_{\rm S}\right)$, commonly employed in non-trace conserving traditional methods. 

Here's why our proposed U-aware HC constraint approach stands out:
\begin{itemize}
    \item \textbf{Generality:} The U-aware HC constraint approach is purely machine learning-based approach and not limited to trace conservation. It can be tailored to enforce various physical constraints across diverse domains within machine learning studies. For example, it could be used to ensure the preservation of total charge in simulations of molecular systems, especially when learning individual charges for each atom.
    \item \textbf{Uncertainty-Aware Correction:} The U-aware HC constraint approach goes beyond simple normalization by incorporating an UQ metric (Eq.~\eqref{eq:uq_metric}) along with a weighting factor (Eq.~\eqref{eq:weight}). This allows for targeted corrections. States (or sites) with greater uncertainty (deviations) receive larger corrections, while those with smaller deviations receive smaller adjustments. This ensures a refined correction process tailored to the level of uncertainty observed.
\end{itemize}

\section{Results and Discussion} \label{sec:results}

In this section, we evaluate the effectiveness of PINNs and our proposed U-aware HC constraint in enforcing trace conservation during simulations. We compare their performance against state-of-the-art purely data-driven neural networks commonly used in open quantum system simulations. For comprehensive assessment, we utilize two distinct processes as benchmarks: relaxation dynamics within the spin-boson (SB) model and the energy transfer process (EET) within the 7-site FMO complex.

For the SB model, we acquire high-quality training data from the publicly available \texttt{QD3SET-1} database.\cite{ullah2023qd3set}  This comprehensive database provides pre-computed dynamics using the hierarchical equations of motion (HEOM) approach.\cite{tanimura1989time, shi2009efficient, chen2022universal} The specific training dataset, denoted by $\mathcal{D}_{\text{sb}}$, consists of 1,000 trajectories simulated across a four-dimensional parameter space encompassing system-bath coupling strength, bath reorganization energy, bath relaxation rate, and inverse temperature (represented by $\epsilon/\Delta$, $\lambda/\Delta$, $\gamma/\Delta$, and $\beta\Delta$, respectively).
```
In similar manner, training data for 7-site FMO complex was also extracted from \texttt{QD3SET-1} database. This dataset encompasses 1,000 training instances, capturing the dynamics for both possible initial excitation sites (site-1 and site-6) within the complex. In the considered data set, the dynamics is propagated for a range of simulation parameters chosen from a three-dimensional space $\mathcal{D}_{\rm fmo} = (\lambda, \gamma, T)$. The method used for propagation is the trace conserving local thermalizing Lindblad master equation (LTLME)\cite{mohseni2008environment} with the system Hamiltonian parameterized by Adolphs and Renger.\cite{adolphs2006proteins} 

\begin{align}
     H_{\rm S}   &=  \nl & \begin{pmatrix}
    200 & -87.7 & 5.5 & -5.9 & 6.7 & -13.7 & -9.9 \\
    -87.7 & 320 & 30.8 & 8.2 & 0.7 & 11.8 & 4.3 \\
    5.5 & 30.8 & 0 & -53.5 & -2.2 & -9.6 & 6.0 \\
    -5.9 & 8.2 & -53.5 & 110 & -70.7 & -17.0 & -63.6 \\
    6.7 & 0.7 & -2.2 & -70.7 & 270 & 81.1 & -1.3\\
    -13.7 & 11.8 & -9.6 & -17.0 & 81.1 & 420 & 39.7 \\
    -9.9 & 4.3 & 6.0 & -63.3 & -1.3 & 39.7 & 230 
    \end{pmatrix}, \label{eq:fmo1-1}
\end{align}
where the diagonal offset is 12210 cm$^{-1}$. 

\begin{figure}
    \centering
    \includegraphics[width=0.5\textwidth]{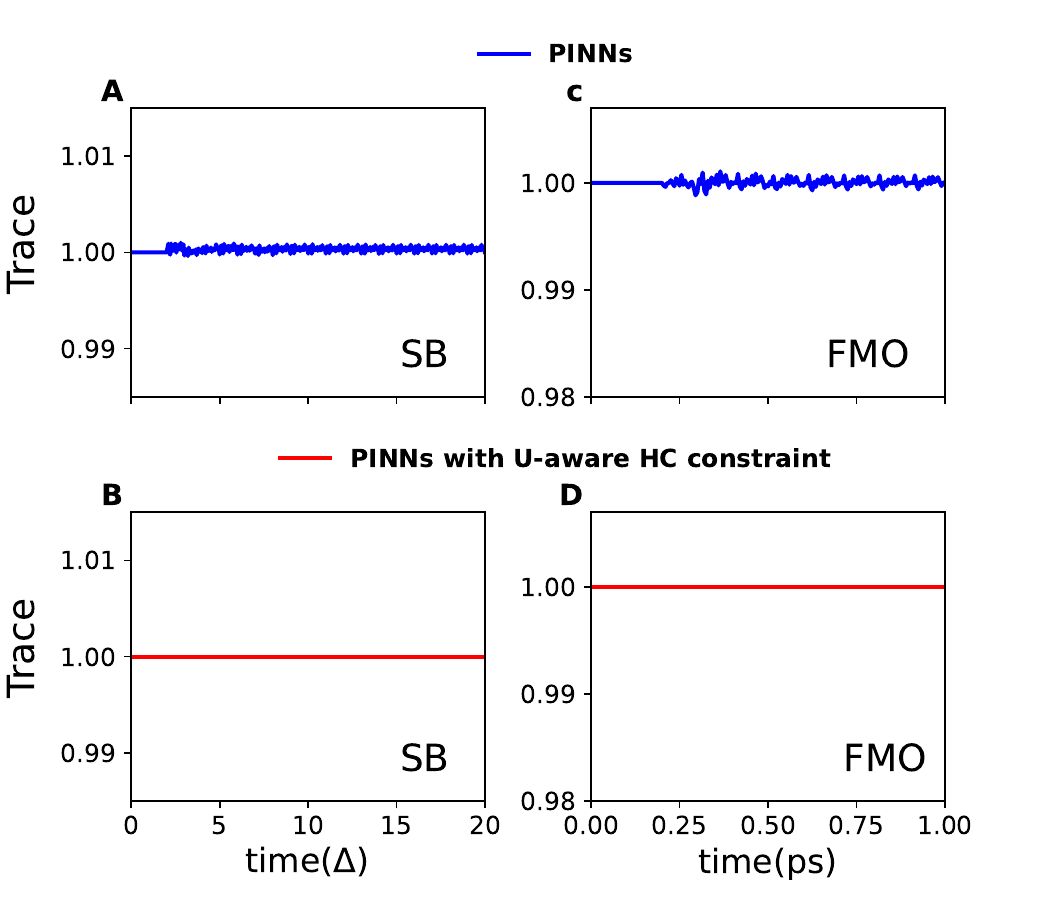}
    \caption{Trace conservation in NN-based simulations using PINNs and the uncertainty-aware HC constraint approach. This figure replicates Fig.~\ref{fig:no-trace-conser} (data-driven NN) for SB model and FMO complex, demonstrating improved conservation with PINNs (Panels~A and C) and perfect trace conservation achieved by combining U-aware HC constraint with PINNs (Panels~B and D). In the case of SB model, an initial period of $t_m\Delta=2.0$ serves as a seed for the model's predictions and results are presented for a test trajectory with characteristic frequency $\gamma/\Delta = 9.0$, system-bath coupling $\lambda/\Delta=0.6$, and inverse temperature $\beta\Delta=1.0$.  For the FMO complex, the initial excitation is considered on site-1, with parameters \(\gamma = 400~\text{cm}^{-1}\), \(\lambda = 40~\text{cm}^{-1}\), and temperature \(T = 90~\text{K}\).}
    \label{fig:hc_trace}
\end{figure}

For the training process, we adopted OSTL-based recursive dynamics propagation (Eq.~\eqref{eq:ostl-rec}) where the RDM $\tilde{\rho}_{\rm s}(t)$ at each time step transforms into a 1D vector with dimension $M$ =  number of sites + (2 $\times$ number of the upper off-diagonal terms). As in RDM $\tilde{\rho}_{\rm S_{nm}}(t) = \tilde{\rho}_{\rm S_{mn}}^*(t) \, (n \neq m)$, only the upper off-diagonal terms are learned. In addition, the real and imaginary parts of each off-diagonal term are separated. More details can be found in Ref.\citenum{ullah2024mlqd}. The target is multi-time step dynamics which is in the same shape as the input. Here we predict the dynamics of 20 time-steps in one shot and which is then fed to the model recursively for the prediction of the next 20 time-steps dynamics. In all cases, we trained a CNN model, implemented in the \texttt{MLQD} package.\cite{ullah2024mlqd} and the uncertainty-aware HC constraint is integrated with priors set as $(p_1, p_2) = (0, 0.1)$.

For the training process, we employed an OSTL-based recursive dynamics propagation approach (Eq.~\eqref{eq:ostl-rec}), where the reduced density matrix (RDM) \(\tilde{\rho}^{(i)}_{\rm s}(t)\) at each time step is transformed into a vector with dimension \(M\), representing the number of sites plus twice the number of upper off-diagonal terms. Since \(\tilde{\rho}_{\rm s_{nm}}(t) = \tilde{\rho}_{\rm s_{mn}}^*(t)\) for \(n \neq m\), only the upper off-diagonal terms are learned, with the real and imaginary parts treated separately. For further details, please refer to Ref.\citenum{ullah2024mlqd}. The target is multi-time step dynamics, maintaining the same shape as the input. We predict the dynamics for 20 time steps in one shot, and these predictions are then recursively fed into the model to forecast the next 20 time steps. Additionally, in our calculations, the uncertainty-aware HC constraint was integrated with priors set at \((p_1, p_2) = (0, 0.1)\), and the \texttt{MLQD} package\cite{ullah2024mlqd} was used for all computations.

To improve training efficiency, we utilized farthest point sampling\cite{dral2019mlatom, ullah2022predicting} to select a subset of training trajectories. For both the symmetric SB model ($\epsilon/\Delta = 0$) and FMO complex with initial excitation on site-1, 400 trajectories were chosen for training, with the remaining used for testing.

In our study, we trained CNN models with identical architectures across all four scenarios. The models used for dynamics propagation yielded nearly identical validation losses, with approximately \(1.2 \times 10^{-5}\) in the SB case and \(1.1 \times 10^{-7}\) in the FMO complex. Introducing trace constraints and adding a prior do impact computational efficiency. Including a trace constraint in the loss function increases its complexity, and the addition of a prior makes the model more challenging to fit, potentially leading to longer training times.

For example, in our experiments, the unconstrained NN model for FMO complex reached a validation loss of \(1.01 \times 10^{-7}\) at epoch 194. In contrast, the PINN model with the same architecture achieved a similar loss of \(1.62 \times 10^{-7}\) at epoch 785, and the auxiliary model in the case of PINN with U-aware HC attained loss of \(2.22 \times 10^{-7}\) at epoch 1142. On our machine (GeForce RTX Nvidia 4090 GPU), each epoch took approximately 1 second, resulting in total training times of 194 seconds, 785 seconds, and 1142 seconds, respectively. While the addition of constraints and priors increases computational time, the overall increase is not significant given the advanced computational resources available today.

Figure~\ref{fig:hc_trace} demonstrates the effectiveness of the PINNs and the Uncertainty-aware HC constraint in maintaining trace conservation during simulations of quantum dissipative dynamics for the SB model and the FMO complex. We revisit the same cases as presented in Fig.~\ref{fig:no-trace-conser} for purely data-driven NNs. As expected, the PINNs (Figs.~\ref{fig:hc_trace} A and C) shows a significant improvement in trace conservation compared to purely data-driven neural networks (Fig.~\ref{fig:no-trace-conser}). However, as previously discussed, PINNs rely on "soft constraints" within the loss function, which can lead to minor deviations from perfect trace conservation.

Perfect trace conservation is achieved via the U-aware HC constraint, as demonstrated in Figs.~\ref{fig:hc_trace} B and D. By explicitly incorporating this constraint within the PINNs framework, we maintain perfect trace conservation throughout the simulations for both the SB model and the FMO complex. This finding underscores the benefit of enforcing strict physical constraints by design, rather than solely relying on the model's ability to learn physical principles indirectly.

Additionally, we present the corresponding population dynamics for all four cases in Fig.~S2 and Fig.~S3 of the Supporting Information. To evaluate the accuracy of each model in dynamics propagation, we provide the MAE averaged over all time steps for each state (site) in Table~I. From the MAE comparison, we observe that all models have tiny errors for populations, so the trace conservation did not have much impact on the quality of the dynamics in the studied cases. However, the trace conservation might have a big impact in the cases where ML struggles to learn and predict dynamics with such an accuracy. As described above, the additional computational cost for enforcing the trace conservation is not that high either, which does not justify the use of the non-conserving approaches in case they break down and have even worse behavior than in Fig.~\ref{fig:no-trace-conser}. In any case, using trace-conserving approaches can be considered as a good prophylactic against unphysical behavior.

\begin{table*}
\caption{MAE averaged over all time-steps (0.2---1~ps) for the diagonal elements \(\tilde{\boldsymbol{\rho}}_{\rm S_{nn}}\) in the SB model and FMO complex. In the table, \(n\) denotes the state (site) number, and Avg(n) represents the average MAE across all states (sites).}
 \begin{tabular}{|l|*{12}{c|}} 
\hline\hline
 &  \multicolumn{2}{c}{SB model} &  & \multicolumn{8}{c}{FMO complex} &  \\  \cline{1-13}
n & \multicolumn{1}{c}{1} & \multicolumn{1}{c}{2} & \multicolumn{1}{c}{Avg(n)}  & \multicolumn{1}{c}{1} & \multicolumn{1}{c}{2}  & \multicolumn{1}{c}{3} &  \multicolumn{1}{c}{4} & \multicolumn{1}{c}{5} &  \multicolumn{1}{c}{6} & \multicolumn{1}{c}{7} & \multicolumn{1}{c}{Avg(n)}\\ \hline
Physics-agnostic NN & 4.37 & 4.20 & 4.29 &  2.57 & 1.39 & 7.91 &  1.99 & 2.06 & 7.62* & 9.12*  & 2.51 \\  
Unconstrained NN &  7.10 & 7.42 & 7.26 & 1.76 &  2.60 &  2.34 & 6.23* & 5.91* &  6.94* & 4.40* & 1.29\\
PINN &  6.14 &  6.10 & 6.12 & 3.93 &  1.13 & 2.76 & 2.15 & 1.55 &  6.15* & 2.17 & 2.04\\
PINN + U-ware HC & 6.08 & 6.08 & 6.08 & 1.41 &  1.53 &  5.22 & 1.26 & 1.54 &  6.73* & 2.10 & 1.96 \\
\hline\hline
\end{tabular}
\vspace{5pt}\\
\centering  All values are in units of $10^{-3}$ except for those marked with *, which are in units of $10^{-4}$\\
\label{tab:ostl}
\end{table*}

.......................................................... 

\section{Concluding remarks} \label{sec:conclusion}

This work addresses the critical issue of trace conservation in NN-based simulations of open quantum systems. While NN models are adept at capturing complex dynamics, they often struggle to maintain fundamental physical principles such as trace conservation. Our investigation reveals three key findings. 

First, purely data-driven NN models, including physics-agnostic and unconstrained NNs, can effectively capture correlations between state-specific populations. However, they lack explicit enforcement of physical laws, leading to potential violations of trace conservation. 

Second, PINNs offer a significant improvement by incorporating physical knowledge into the loss function. This method penalizes deviations from physical constraints, enhancing the accuracy of simulations. Despite this advancement, PINNs still rely on "soft constraints," which can result in minor violations of physical constraints like trace conservation. 

Finally, U-aware HC constraint approach addresses the limitations of PINNs by enforcing trace conservation by design rather than solely through the loss function. The U-aware HC constraint utilizes uncertainty quantification techniques to redistribute residual errors and correct potential trace violations, ensuring physically consistent simulations throughout. 

It is important to note that while we did not explicitly enforce a positivity constraint in our case--since all diagonal elements remained strictly positive--such a constraint could be incorporated if necessary.

To conclude, our findings underscore the importance of integrating well-defined physical constraints into NN models. The methods developed in this study are broadly applicable and can be adapted to enforce other essential constraints in various domains. For instance, in molecular simulations where individual atomic charges are learned, our different-prior approach for uncertainty quantification as well as an approach for redistributing residual error in atomic charges could be used as an alternative to existing, related approaches~\cite{charge1,charge2} for ensuring total charge conservation. By extending these techniques, we can improve the fidelity and reliability of NN-based simulations across a wide range of scientific and engineering applications.


\section{Acknowledgments}

A.U. acknowledges funding from the National Natural Science Foundation of China (No. W2433037). P.O.D. acknowledges support from the National Natural Science Foundation of China (No. 22003051), as well as funding through the Outstanding Youth Scholars (Overseas, 2021) project and the Fundamental Research Funds for the Central Universities (No. 20720210092). This project is also supported by the Science and Technology Projects of the Innovation Laboratory for Sciences and Technologies of Energy Materials of Fujian Province (IKKEM) (No. RD2022070103).

\section{Data availability}
The data supporting this work is available at \url{https://github.com/Arif-PhyChem/trace_conservation}.

\section{Competing interests}
The authors declare no competing interests.

\section*{References}
\bibliography{main.bib}

\end{document}


\title{Supporting Information for "Physics-Informed Neural Networks and Beyond: Enforcing Physical Constraints in Quantum Dissipative Dynamics"}

\author{Arif Ullah}   
\email{arif@ahu.edu.cn}
\author{Yu Huang}   
\author{Ming Yang}
\affiliation{School of Physics and Optoelectronic Engineering, Anhui University, Hefei, 230601, Anhui, China}  
\author{Pavlo O. Dral}
\email{dral@xmu.edu.cn}
\affiliation{State Key Laboratory of Physical Chemistry of Solid Surfaces, College of Chemistry and Chemical Engineering, Fujian Provincial Key Laboratory of Theoretical and Computational Chemistry, and Innovation Laboratory for Sciences and Technologies of Energy Materials of Fujian Province (IKKEM), Xiamen University, Xiamen, 361005, Fujian, China}
\affiliation{Institute of Physics, Faculty of Physics, Astronomy, and Informatics, Nicolaus Copernicus University in Toruń, ul. Grudziądzka 5, 87-100 Toruń, Poland}

\date{\today}

\begin{abstract}
\end{abstract}

\maketitle

\begin{suppfigure}
     \centering
     \includegraphics[width=\textwidth]{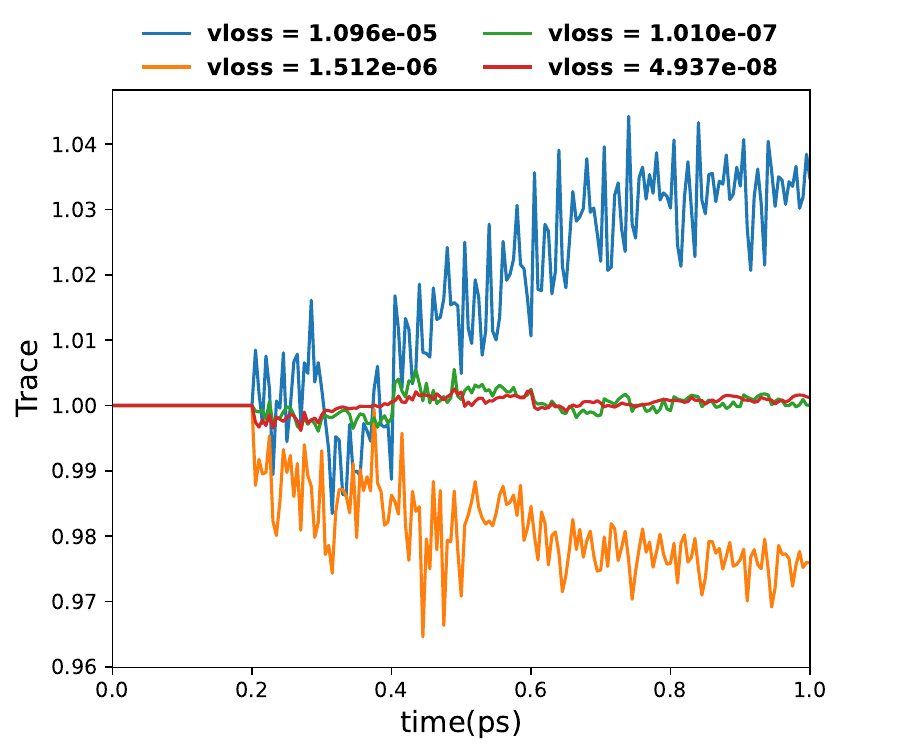}
     \caption{Comparison of trace conservation for unconstrained NN with varying validation loss (vloss). The considered system is the 7-site FMO complex, where an initial dynamics of 0.2 ps, exhibiting ideal trace conservation, is used as the seed for model predictions based on reference calculations. The initial excitation is located on site-1, with parameters \(\gamma = 400~\text{cm}^{-1}\), \(\lambda = 40~\text{cm}^{-1}\), and temperature \(T = 90~\text{K}\). Further details on training and prediction are provided in the Results and Discussion section of the main text.}
     \label{fig:sb_dyn}
 \end{suppfigure}

\begin{supptable}
\caption{Mean Absolute Error (MAE) averaged over time steps from 0.8 to 1~ps for the diagonal elements \(\tilde{\boldsymbol{\rho}}_{\rm S_{nn}}\) in the FMO complex, considering different memory times, denoted as $t_m$. In the table, $n$ represents the state site number, and Avg(n) denotes the average MAE across all sites. Additionally, the MAE for trace is calculated and averaged over all time steps within the same time interval (0.8 to 1~ps). We used unconstrained NNs where all the models had nearly identical validation mean square error (loss), approximately ~$1.0 \times 10^{-7}$.}
 \begin{tabular}{|l|*{10}{c|}} 
\hline\hline
 &  \multicolumn{8}{c}{Average MAE for each site $n$} & \multicolumn{1}{c}{MAE for Trace} \\  \cline{1-10}
n & \multicolumn{1}{c}{1} & \multicolumn{1}{c}{2}  & \multicolumn{1}{c}{3} &  \multicolumn{1}{c}{4} & \multicolumn{1}{c}{5} &  \multicolumn{1}{c}{6} & \multicolumn{1}{c}{7} & \multicolumn{1}{c}{Avg(n)} &  \\ \hline
$t_m = 0.2$~ps & 1.50e-3 & 1.0e-1 & 1.30e-1 &  1.05e-1 & 9.42e-2 & 3.10e-2 &  1.22e-2 & 6.80e-2 &  7.31e-4 \\  
$t_m = 0.4$~ps &  9.77e-3 & 2.15e-3 & 1.12e-2 & 1.96e-3 &  1.23e-3 &  1.92e-3 & 9.44e-4 &  4.18e-3 & 2.79e-3\\
$t_m = 0.6$~ps &  8.10e-4 &  5.01e-3 & 1.19e-3 & 5.46e-4 &  1.35e-3 & 1.48e-3 & 1.46e-3 & 1.16e-3 & 1.17e-3\\
$t_m = 0.8$~ps & 3.27e-3 & 2.92e-3 & 2.49e-3 & 1.34e-3 &  2.59e-4 &  6.04e-4 & 5.75e-4 & 1.63e-3 & 9.10e-4 \\
\hline\hline
\end{tabular}
\vspace{5pt}
\label{tab:ostl}
\end{supptable}

\begin{suppfigure}
     \centering
     \includegraphics[width=\textwidth]{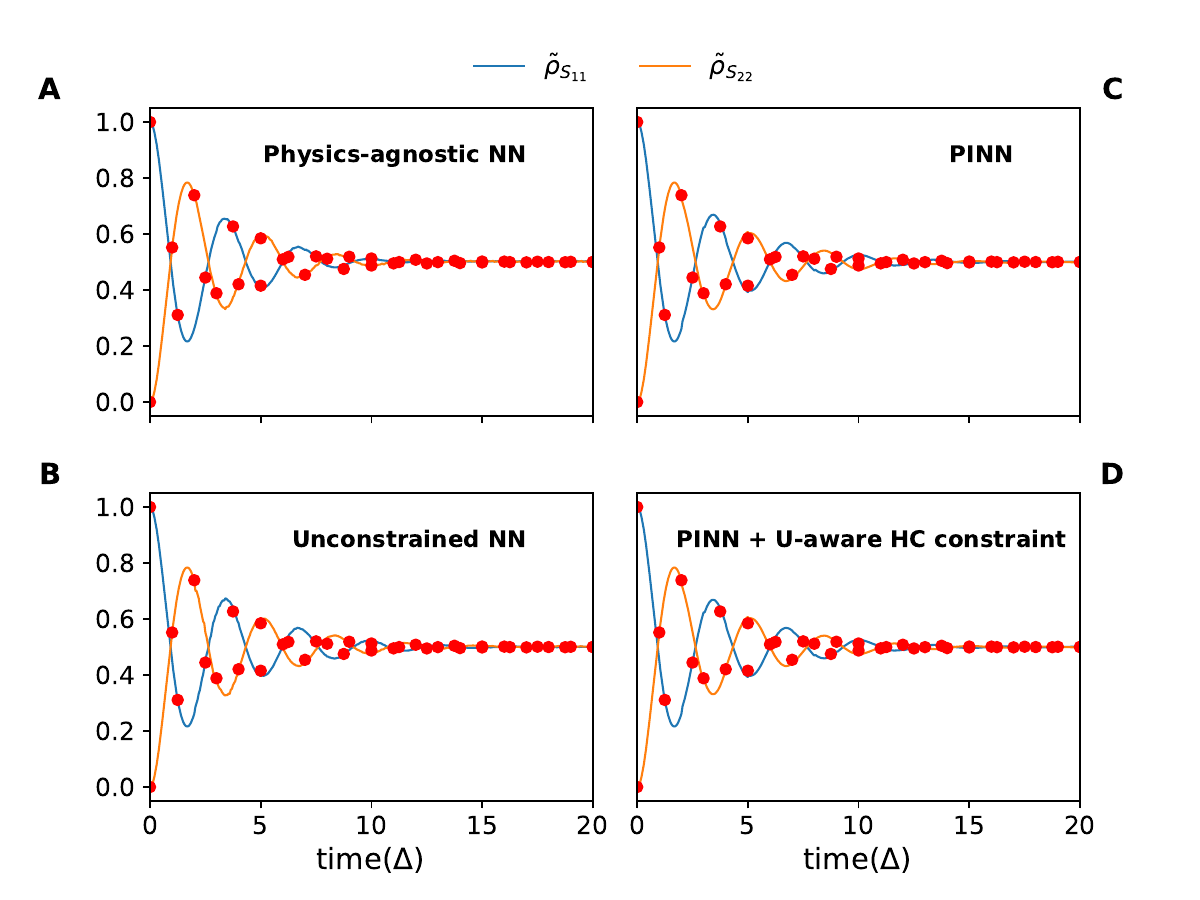}
     \caption{Population dynamics of the two states in the symmetric SB model as a function of time. Results are presented for an unseen trajectory with $\gamma/\Delta = 9.0$, $\lambda/\Delta = 0.6$, and $\beta\Delta = 1.0$. A short HEOM dynamics with a time length of $t_m\Delta = 2.0$ was used as a seed and recursive dynamics was propagated with 20 time steps in one shot. The results are compared with HEOM results (dots).}
     \label{fig:sb_dyn}
 \end{suppfigure}

 \begin{suppfigure}
     \centering
     \includegraphics[width=\textwidth]{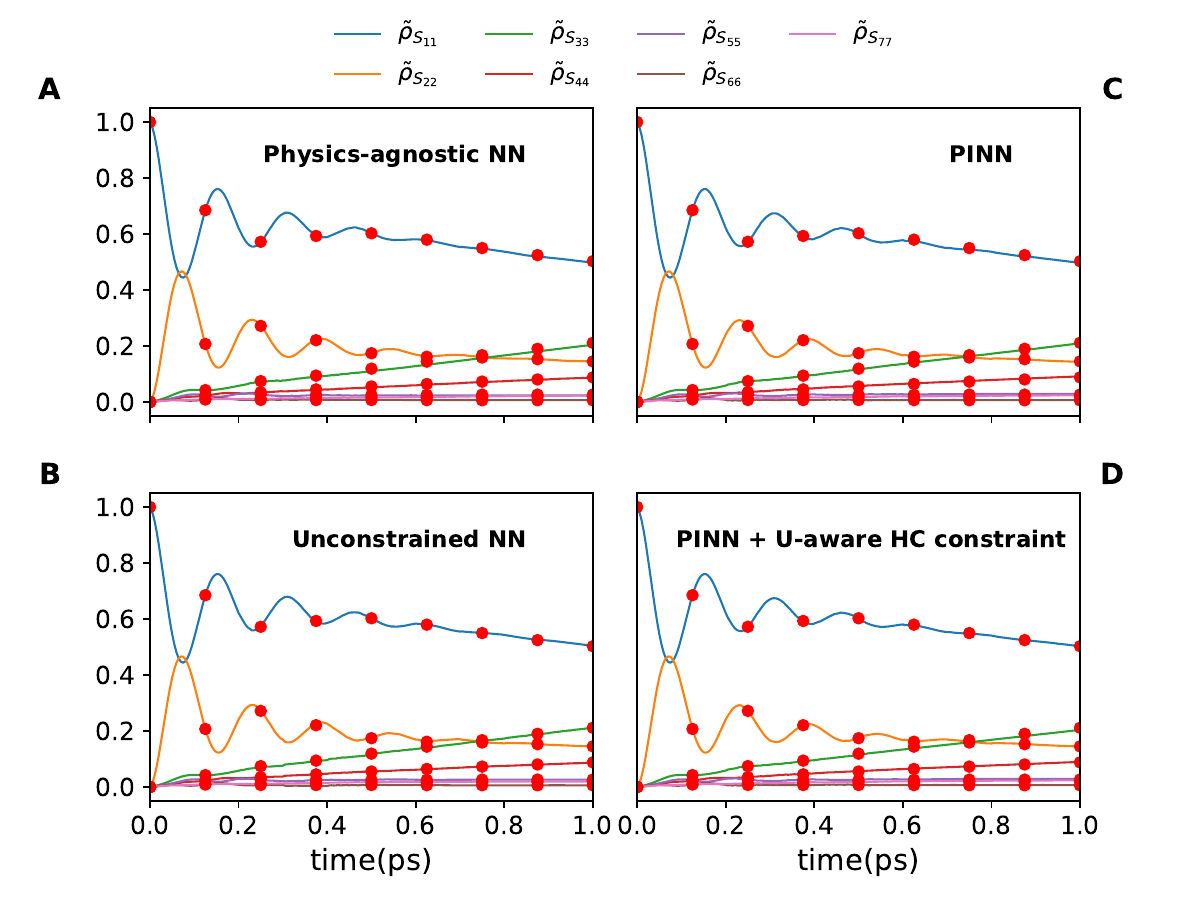}
     \caption{Excitation energy transfer in the 7-site FMO complex as a function of time. Results are presented for an unseen trajectory with $\gamma = 400.0~\text{cm}^{-1}$, $\lambda = 40.0~\text{cm}^{-1}$, and $T = 90.0~\text{K}$. A short LTLME dynamics with a time length of $t_m\Delta = 0.2$~ps was used as a seed and recursive dynamics was propagated with 20 time steps in one shot. The results are compared with LTLME results (dots).}
     \label{fig:sb_dyn}
 \end{suppfigure}